\documentclass[a4paper,11pt]{article}
\pdfoutput=1 

\usepackage{jinstpub} 

\usepackage[nooneline]{subfigure}
\subfiguretopcaptrue

\usepackage[separate-uncertainty=true]{siunitx}

\title{Multimodal Shifts in the Dark Current of Silicon Photomultipliers}

\author[a]{Ryu~Kawarasaki,}
\author[a,b,c,1]{Akira~Okumura,\note{Corresponding author.}}
\author[a]{Kazuhiro~Furuta,}
\author[a,b,c]{and Hiroyasu~Tajima}
\author{for~the CTAO Consortium}

\affiliation[a]{Institute for Space--Earth Environmental Research, Nagoya University,\\Furo-cho, Chikusa-ku, Nagoya 464-8601, Japan}
\affiliation[b]{Kobayashi--Maskawa Institute for the Origin of Particles and the Universe, Nagoya University\\Furo-cho, Chikusa-ku, Nagoya 464-8602, Japan}
\affiliation[c]{Nagoya University Southern Observatories, Nagoya University\\Furo-cho, Chikusa-ku, Nagoya 464-8602, Japan}

\emailAdd{oxon@mac.com}

\abstract{The dark count rate is one of the key properties of avalanche photodiodes (APDs) and silicon photomultipliers (SiPMs). Previous studies have reported discrete shifts in the dark count rate on short timescales (\qty{\sim 10}{\ms} to \qty{\sim 100}{\s}) with small increases (up to \qty{\sim 1}{\kHz} per APD). In this study, we report a similar yet distinct phenomenon in the dark current of SiPMs designed for gamma-ray telescopes. Long-term stability tests (${>}100$\,days) revealed bimodal or multimodal dark current shifts of the order of \qty{0.1}{\uA} on timescales of days in 48 out of 128 SiPM channels. In addition, optical emission was observed from an APD surface when the dark current was in a high state. These findings suggest that multimodal dark current behavior is a common property of SiPMs, which may be due to defects in the silicon.}


\proceeding{The 6$^{\text{th}}$ International Workshop on New Photon-Detector (PD24)\\
 Nov 19--22, 2024\\
 Vancouver (BC), Canada}

\begin{document}
\maketitle
\flushbottom

\section{Introduction}
\label{sec:Introduction}


The dark count rate (DCR) of silicon photomultipliers (SiPMs) is one of their key properties that impacts the trigger performance, energy bias, and energy resolution of particle detectors. Since the dark current---proportional to the product of the DCR and the gain---varies with temperature, bias voltage, and avalanche probability, its monitoring is essential for ensuring detector stability. The total dark current from all avalanche photodiode (APD) cells within a single SiPM channel serves as an indicator of system instabilities or malfunctions.


Previous studies have shown that the DCR of a single APD with a small area (${<\mathcal{O}(1000)}$\,\unit{\um\squared}) and that of millimeter-scale SiPMs at low temperature can exhibit discrete fluctuations \cite{Capua:2021:Investigation-of-random-telegraph-signal-in-two-ju,Tsang:2023:Studies-of-event-burst-phenomenon-with-SiPMs-in-li}. In small APDs, these fluctuations are attributed to random telegraph signals (RTSs), which are commonly observed in semiconductor devices due to defects in doped silicon \cite{Ralls:1984:Discrete-Resistance-Switching-in-Submicrometer-Sil}. In contrast, for larger SiPMs at lower temperatures, discrete changes in DCR may be triggered by cosmic-ray muons.

These previously observed discrete DCR fluctuation are relatively short-lived and small in magnitude, with timescales ranging from \qty{\sim 10}{\ms} to \qty{\sim 100}{\s}, and DCR increases of up to \qty{\sim 1}{\kHz} per APD. However, longer timescale or larger-magnitude variations in SiPM dark current have not been well studied.

In this study, we report the discovery of a third type of discrete dark current change, observed serendipitously during long-term dark current monitoring of SiPMs intended for the Small-Sized Telescopes (SSTs) of the Cherenkov Telescope Array Observatory (CTAO) \cite{White:2022:The-Small-Sized-Telescopes-for-the-Southern-Site-o}, a next-generation ground-based very-high-energy gamma-ray observatory. The initial goal of this monitoring was to assess SiPM failure rates and identify any unforeseen issues before mass production of CTAO components. However, we found that at least 48 out of 128 SiPM channels exhibited unexpected dark current behaviors, where the average dark current randomly shifted between discrete levels (bimodal or multimodal patterns) on long timescales of minutes to days, with current shifts $\Delta I$ on the order of \qty{0.1}{\uA} per channel (equivalent to $\mathcal{O}(1)$\,\unit{\MHz} per channel).

Section~\ref{sec:methods} describes the SiPM specifications and the dark current monitoring methodology. Section~\ref{sec:results} presents results from over $100$\,days of data collection. In Section~\ref{sec:discussion}, we discuss the observed multimodal behavior and propose a possible explanation.

\section{Methods}
\label{sec:methods}

\begin{figure}[b]
\centering
\includegraphics[width=.6\textwidth]{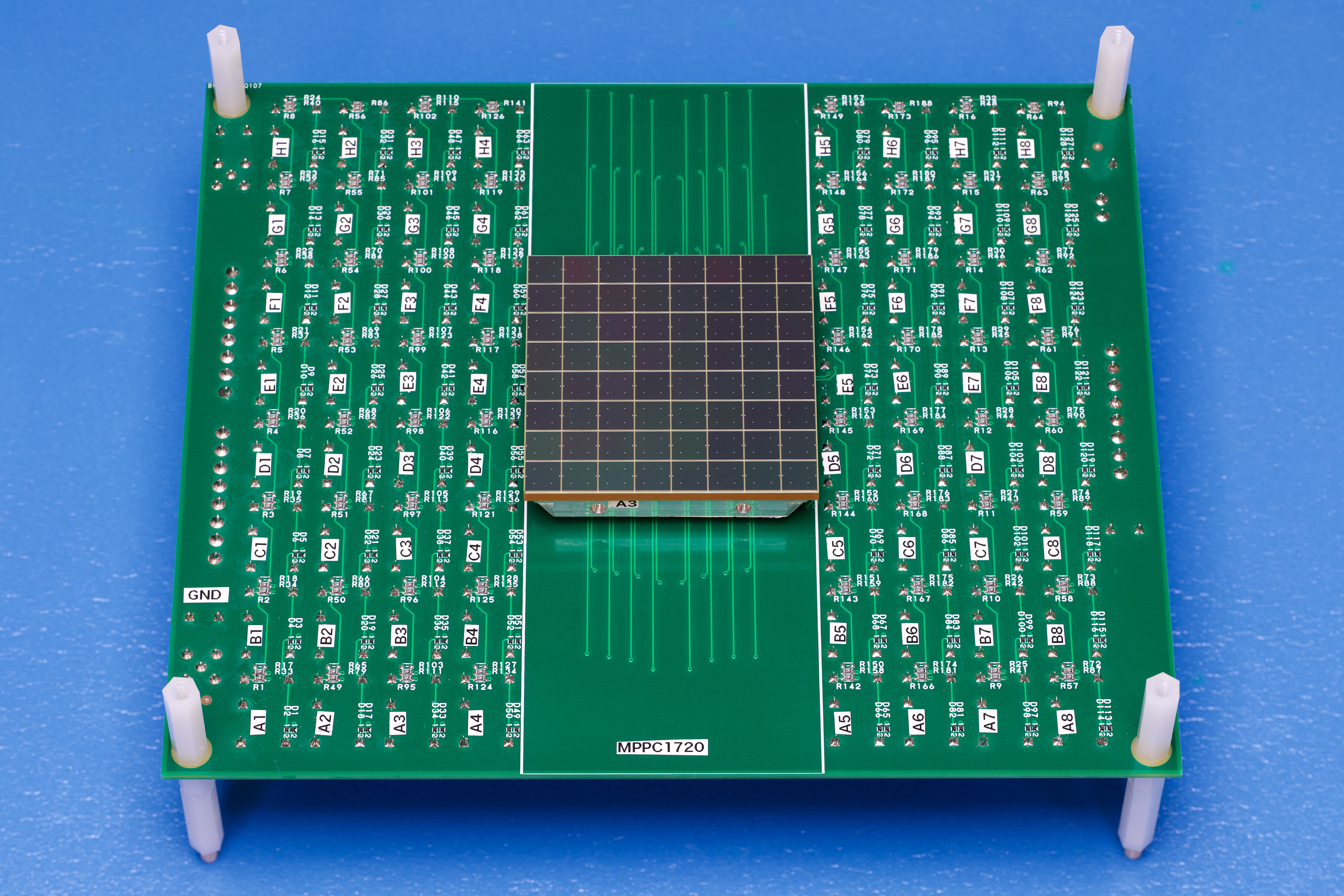}
\caption{Photo of an S14521-1720 module and its readout board, the back of which is equipped with a switching relay array.}
\label{fig:SiPM}
\end{figure}

The focal-plane camera of CTAO SSTs will consist of 2048 SiPM channels and readout electronics \cite{Schwab:2024:CTC-and-CT5TEA:-An-advanced-multi-channel-digitize,White:2022:The-Small-Sized-Telescopes-for-the-Southern-Site-o}, where 32 SiPM modules, each with 64 channels, are spherically aligned to cover the \qty{\sim 9}{\degree} field of view. S14521-1720 (Hamamatsu Photonics), a 64-ch ultraviolet (UV)-sensitive SiPM module, has been selected for SSTs (Fig.~\ref{fig:SiPM}). Each individual channel of S14521-1720 have a sensitive area of \qtyproduct{6 x 6}{\mm}, populated with \qtyproduct{50x50}{\um} APD cells.

S14521-1720 was specifically customized for Cherenkov telescopes to enhance the photon detection efficiency in the UV waveband ($300$--$400$\,\unit{nm}) and to suppress the optical crosstalk rate. The typical breakdown voltage was designed to be approximately \qty{39}{\V}, and the protective window on the photodiode surface is made of only thin SiO$_2$ and Si$_3$N$_4$ layers.

To study long-term stability and identify potential issues before the construction of the SSTs, we monitored the dark currents of S14521-1720 serial numbers (SN) 9 and 13 (128 channels total) for 110 days and 162 days, respectively. The ambient temperature during the monitoring was maintained at $24.3\pm0.2$\,\unit{\degreeCelsius} by placing the modules in an environmental chamber (Espec LHU-124).

Building a monitor system capable of reading the currents of all 128 channels simultaneously would be unnecessarily expensive, as it would require the same number of source meters to supply bias voltages to individual channels. Instead, our dark current monitoring system reads the voltage across a 100-$\Omega$ resistor connected in series with each SiPM channel. A common bias voltage of \qty{44}{\V} (equivalent overvoltage of \qty{\sim 5}{\V}) was supplied to the 128 channels from a Keithley 2410 source meter via a dedicated readout board (Fig.~\ref{fig:SiPM}). The resistor voltages were read sequentially every several minutes using an onboard array of relay circuits and an external multimeter (Keithley 2700).

In addition to current monitoring, the SiPM surfaces were optically inspected by using a digital camera (Canon EOS 6D) and a macro lens (Canon MP-E 65mm f/2.8 1--5x) to search possible emission from silicon de-excitation.

\section{Results}
\label{sec:results}

\begin{figure}
  \centering
  \subfigure[]{%
    \includegraphics[width=.4\textwidth]{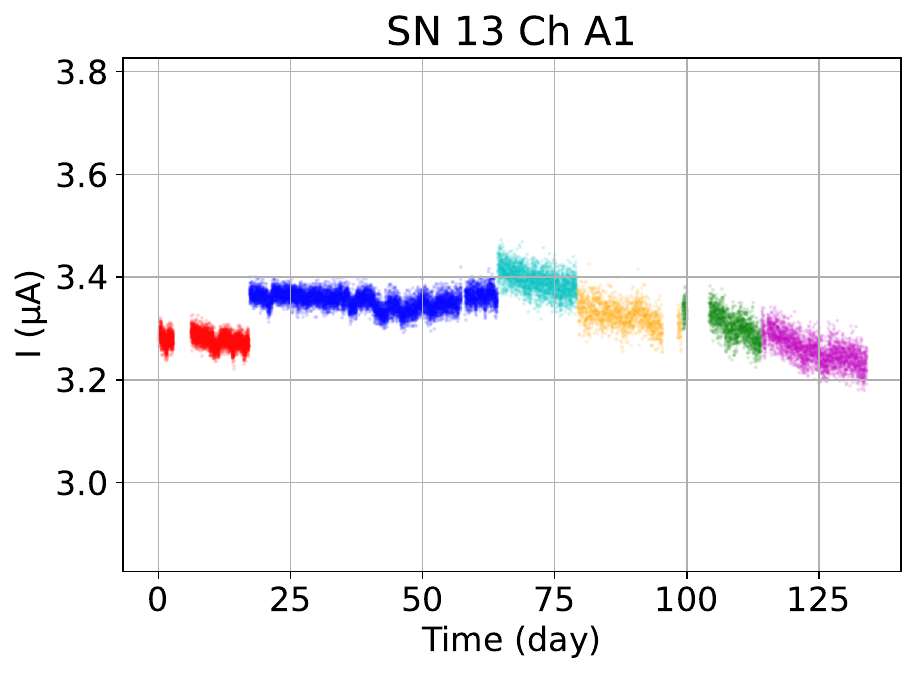}
    \label{fig:multimodal-A1-long}%
  }%
  \subfigure[]{%
    \includegraphics[width=.4\textwidth]{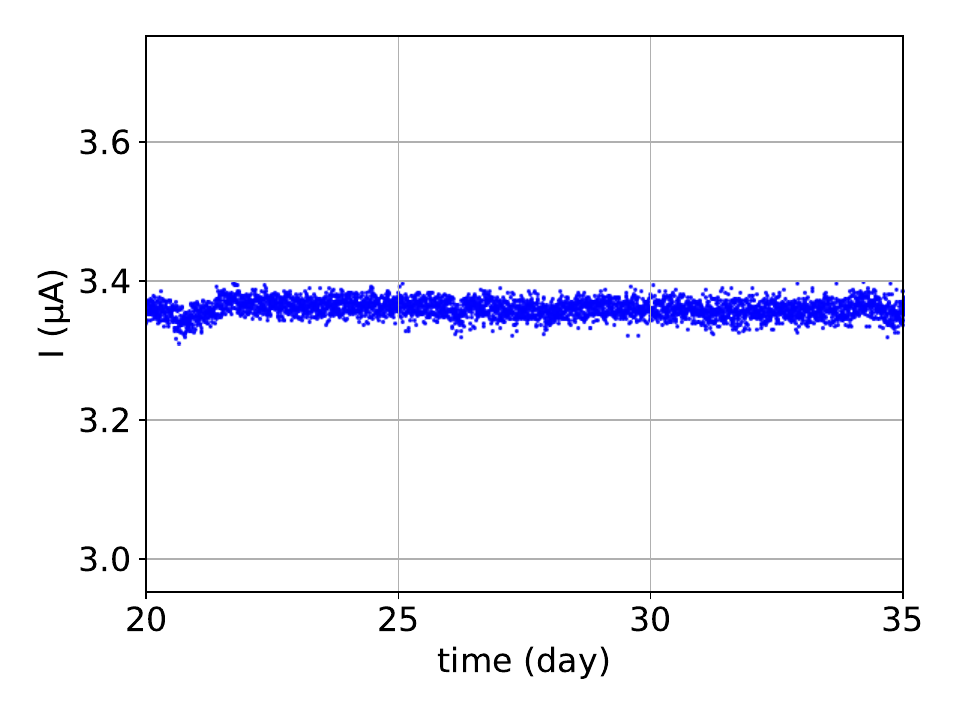}
    \label{fig:multimodal-A1}%
  }
  \subfigure[]{%
    \includegraphics[width=.4\textwidth]{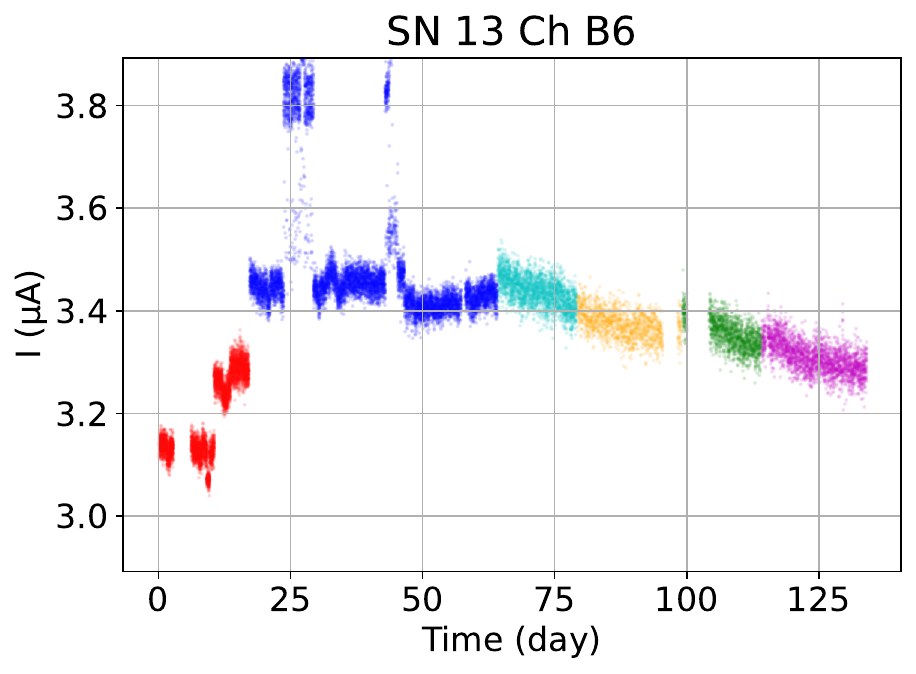}
    \label{fig:multimodal-B6-long}%
  }%
  \subfigure[]{%
    \includegraphics[width=.4\textwidth]{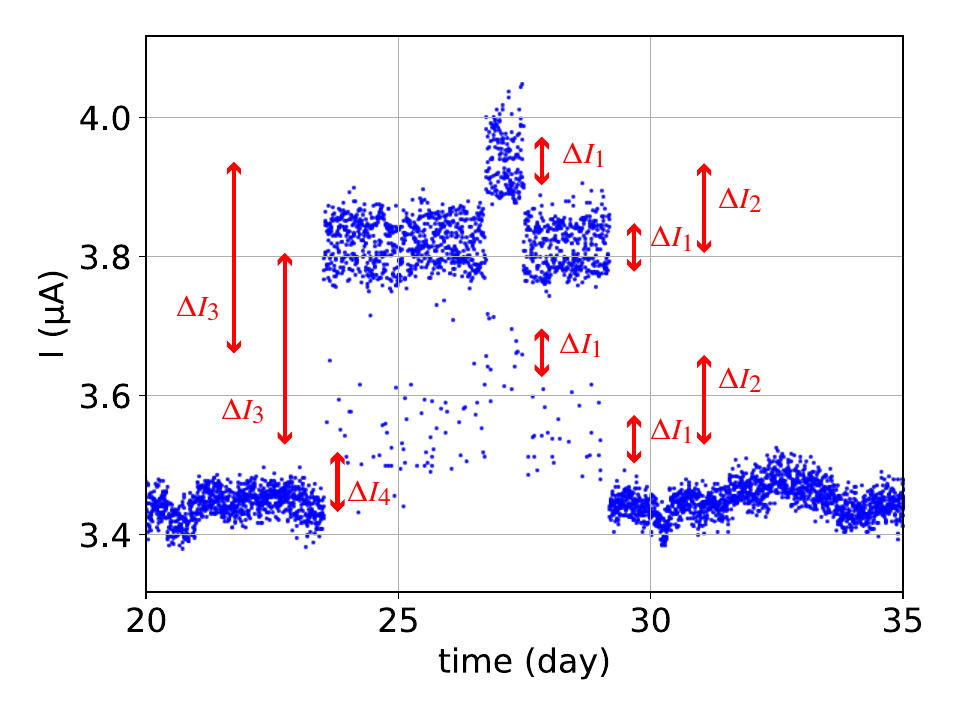}
    \label{fig:multimodal-B6}%
  }
  \subfigure[]{%
    \includegraphics[width=.4\textwidth]{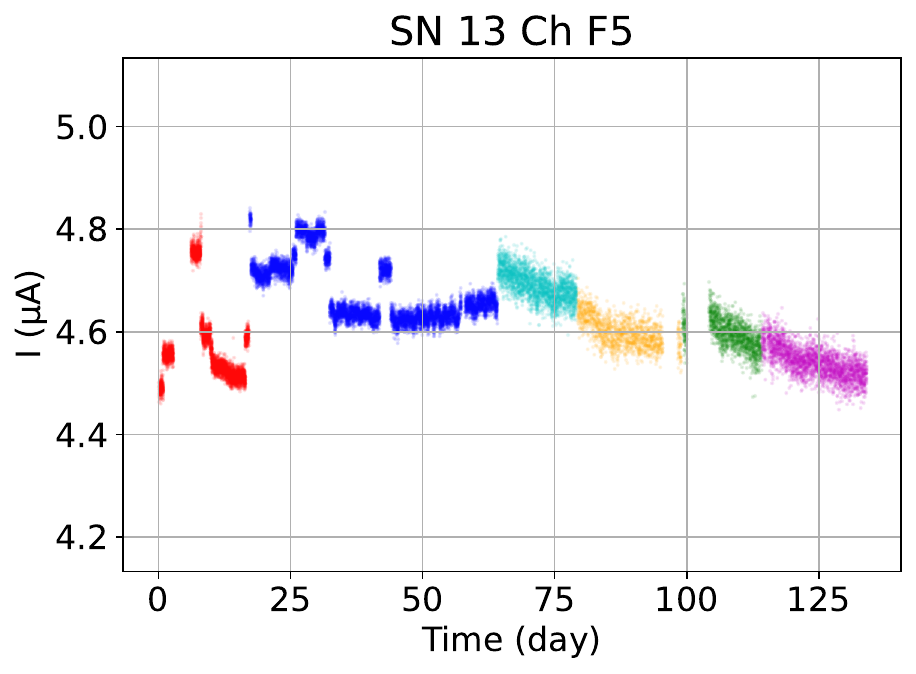}
    \label{fig:multimodal-F5-long}%
  }%
  \subfigure[]{%
    \includegraphics[width=.4\textwidth]{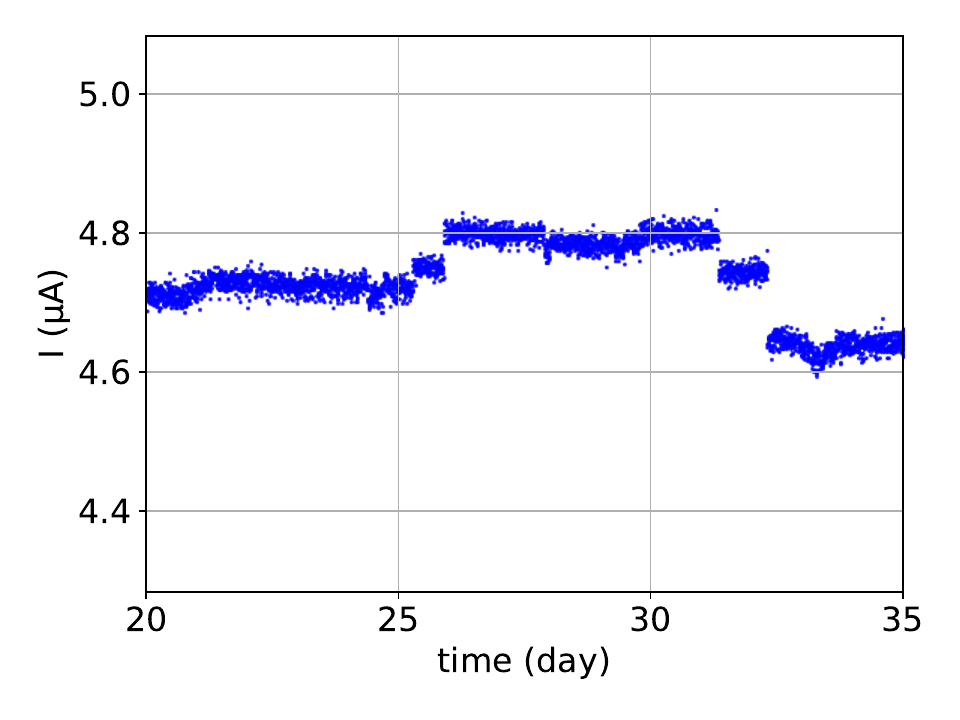}
    \label{fig:multimodal-F5}%
  }
  \subfigure[]{%
    \includegraphics[width=.4\textwidth]{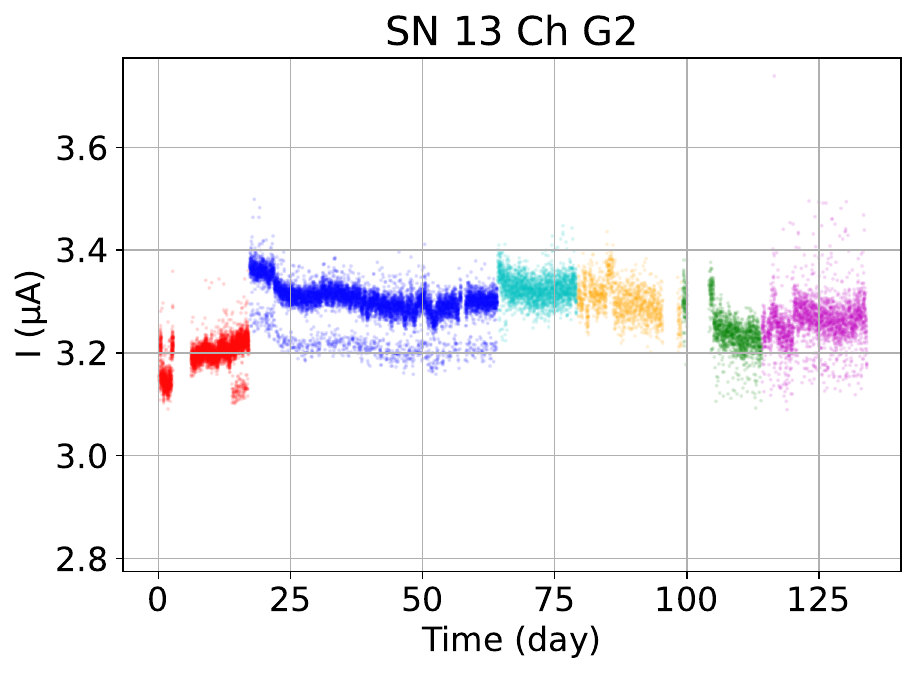}
    \label{fig:multimodal-G2-long}%
  }%
  \subfigure[]{%
    \includegraphics[width=.4\textwidth]{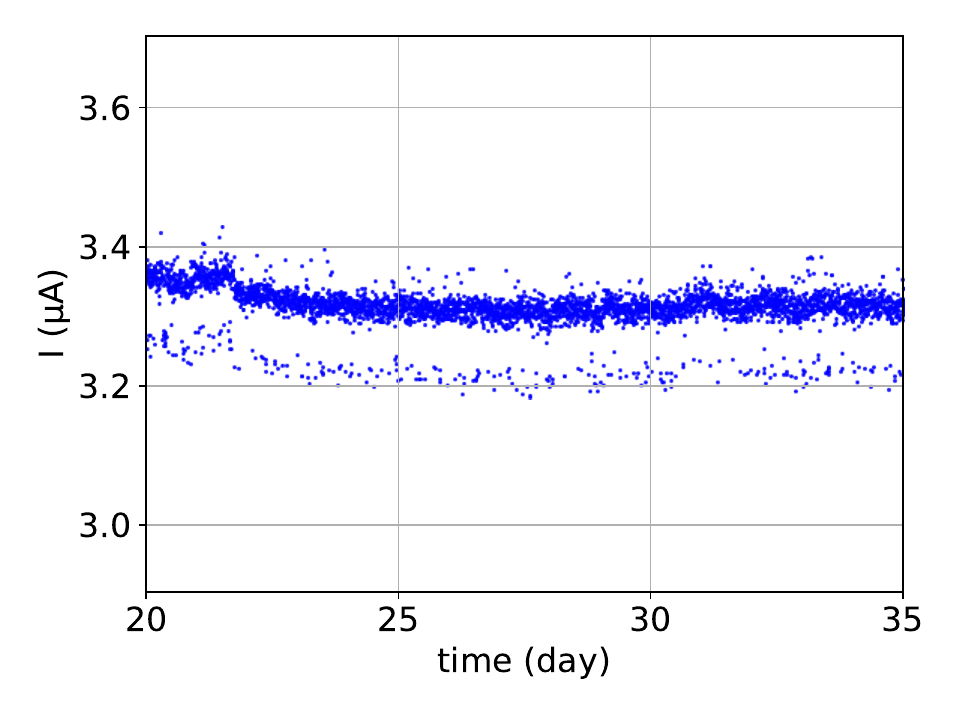}
    \label{fig:multimodal-G2}%
  }%
  \caption{Dark currents of Ch A1 (a), B6 (c), F5 (e), and G2 (g) of SN~13 over the entire monitoring period. The vertical axis ranges differ across plots, but the scale remains consistent. To better highlight discrete shifts, a shorter period from days 20 to 35 is shown in (b), (d), (f), and (h).}
  \label{fig:multimodal}
\end{figure}

Fig.~\ref{fig:multimodal} shows the temporal changes of the dark current of four selected channels of SN~13. Figures~\ref{fig:multimodal-A1-long} and \ref{fig:multimodal-A1} correspond to a representative stable channel, A1. Different colors indicate six system configurations, which introduced systematic measurement offsets of approximately \qty{0.1}{\uA} even in stable channels.

Figures~\ref{fig:multimodal-B6-long}, \ref{fig:multimodal-F5-long}, and \ref{fig:multimodal-G2-long} show examples of unstable channels whose multimodal or bimodal current shifts are clearly visible in Figures~\ref{fig:multimodal-B6}, \ref{fig:multimodal-F5}, and \ref{fig:multimodal-G2}. The fact that these discrete current changes occur asynchronously rules out any instability in the chamber temperature, common bias voltage, or readout system as potential causes. This conclusion is further supported by the observation that both the timescales and $\Delta I$ values remain consistent across the first (red) and second (blue) configuration periods, despite the readout board being replaced. In addition to Chs B6, F5, and G2 of SN~13, 48 out of 128 channels exhibited bimodal and multimodal dark current behavior, identified through visual inspection on the graphs. Therefore, we conclude that multimodal dark current is an intrinsic property of SiPMs, or at least of the S14521-1720 model.

The multimodal behavior seen in Fig.~\ref{fig:multimodal} provides further insight into the underlying mechanism within the SiPMs. First, as indicated by the arrows in Fig.~\ref{fig:multimodal-B6}, the multimodal shifts can be decomposed into sums of distinct current shifts $\Delta I_i$. For instance, the vertically spread band in $3.8$--$4.0$\,\unit{\uA} can be separated into contributions from two different shifts, $\Delta I_1$ and $\Delta I_2$. A similar structure is found in the $3.5$--$3.7$\,\unit{\uA} range, which is shifted downward by $\Delta I_3$, and this band is also $\Delta I_4$ away from the \qty{\sim 3.4}{\uA} baseline. These observations raise a hypothesis that the observed current shifts originate from multiple active components present in the single channel---namely, individual APD cells.

Interestingly, each multimodal channel has different timescales and current differences. The switching speed of Ch G2 (Fig.~\ref{fig:multimodal-G2}) and some components of Ch B6 ($\Delta I_1$ and $\Delta I_3$ in Fig.~\ref{fig:multimodal-B6})) is faster than our readout time resolution of about 5 minutes. However, the other shifts occur on timescales of days.

To test the hypothesis that the transient large dark current states originate from a few noisy APD cells, we continuously captured optical images of the surface of Ch E2 of SN 9, a bimodal channel, with 30 second exposures. The images were recorded in Canon RAW format to preserve the original linearity of the image sensor for analysis. If a dark current increase is caused by very frequent (MHz) avalanche processes in a single APD cell, there should be visible to near-infrared emission at the same location, resulting from silicon de-excitation.

\begin{figure}
  \centering
  \subfigure[]{%
    \includegraphics[width=.4\textwidth]{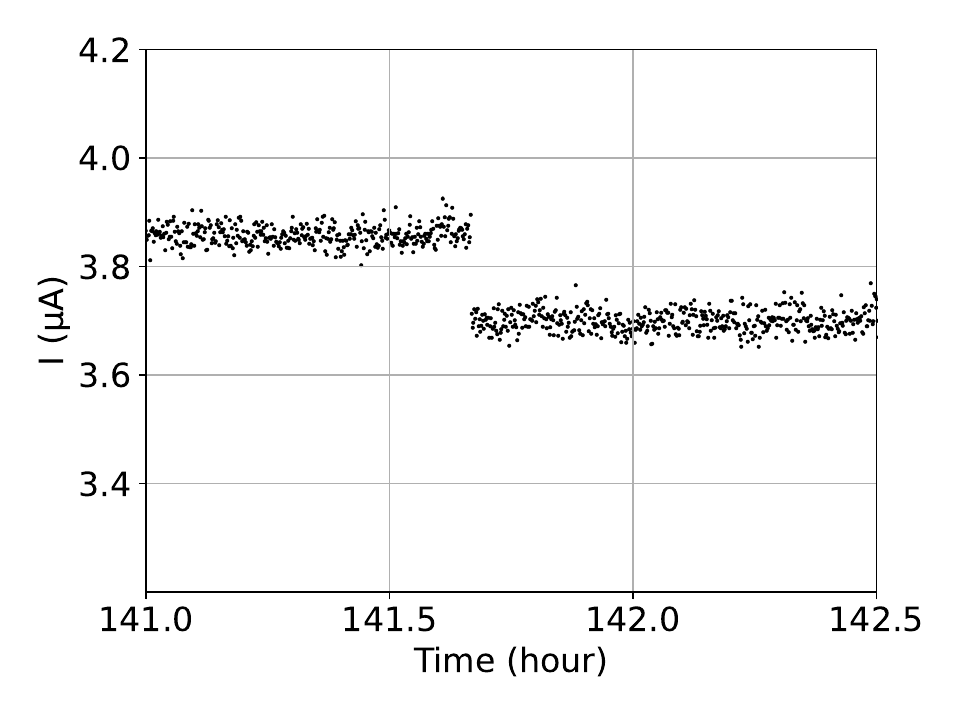}
    \label{fig:SN9-E2}%
  }
  \subfigure[]{%
    \includegraphics[width=.4\textwidth]{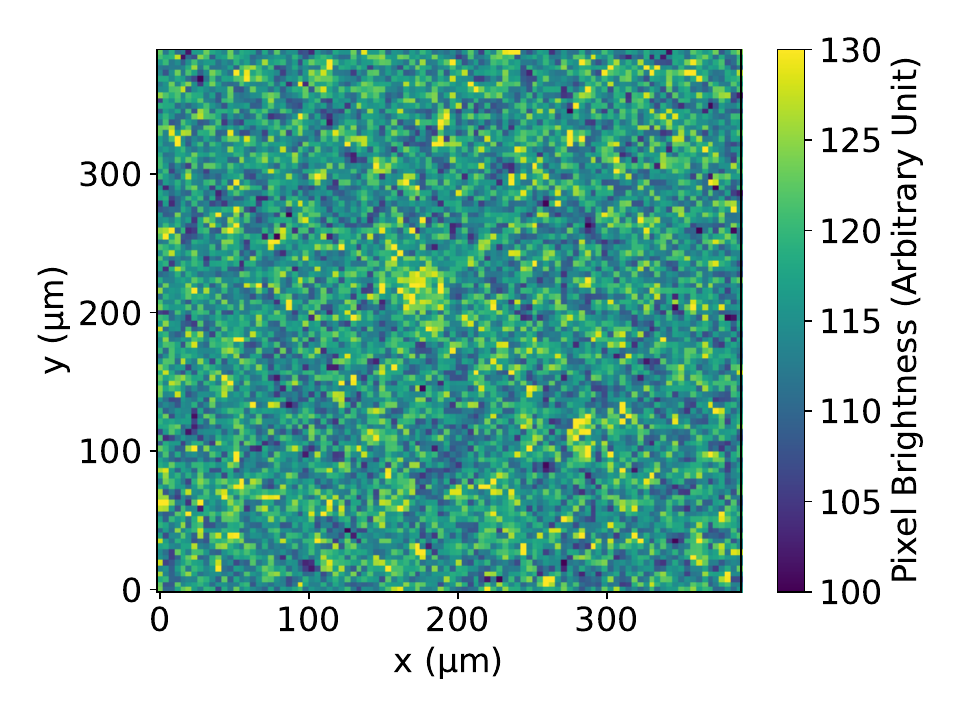}
    \label{fig:SN9-E2-before}%
  }
  \subfigure[]{%
    \includegraphics[width=.4\textwidth]{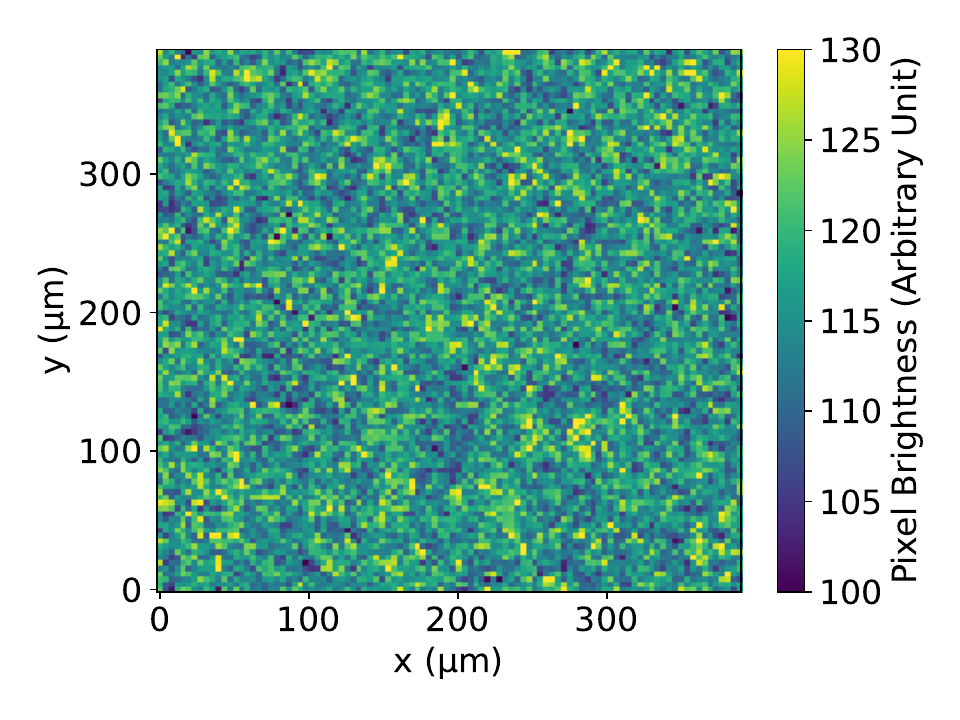}
    \label{fig:SN9-E2-after}%
  }%
  \subfigure[]{%
    \includegraphics[width=.4\textwidth]{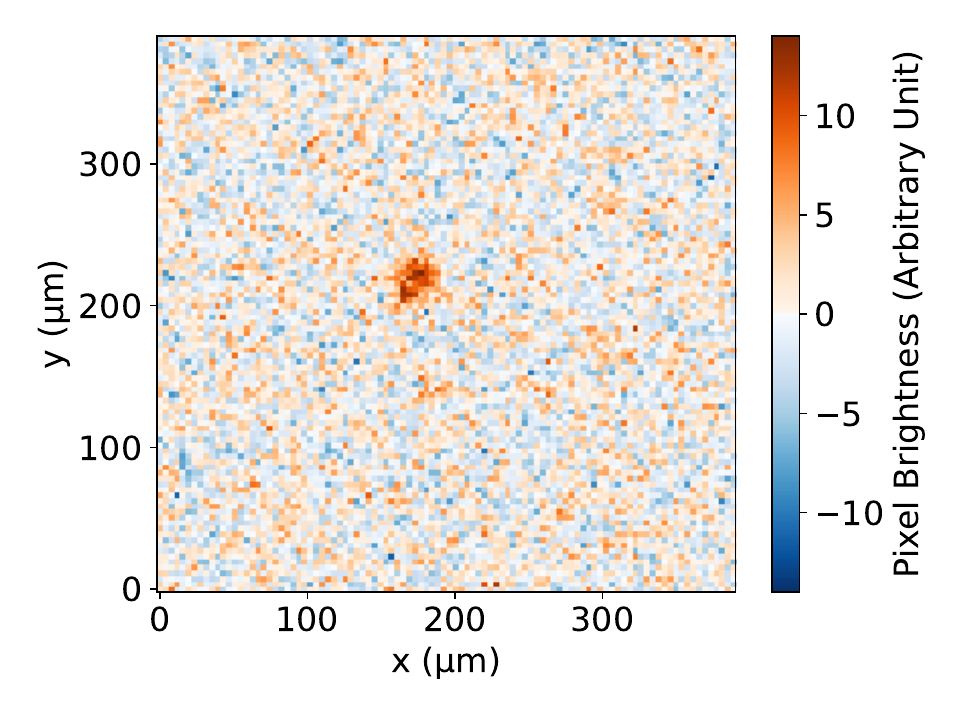}
    \label{fig:SN9-E2-diff}%
  }%
  \caption{(a) Monitored dark current of Ch E2 of SN 9, showing a current drop at 141.7 hours. (b) Stacked surface image of Ch E2 during the last 38 min before the current drop. (c) Stacked surface image of the first 38 min after the current drop. (d) Difference image between (b) and (c), highlighting changes in optical emission. Note that the APD cell is \qty{50}{\um}.}
  \label{fig:hotspot}
\end{figure}

Fig.~\ref{fig:SN9-E2-before} shows the dark current of Ch E2 of SN 9, which was measured approximately every 6 seconds during the continuous imaging of the E2 surface. A current drop from \qty{\sim 3.8}{\uA} to \qty{\sim 3.7}{uA} occurred at $141.7$\,hours, consistent with the bimodal nature of E2.

To improve the signal-to-noise ratio, we stacked multiple images from the last $38$ minutes ($60$ frames, each with a \qty{30}{\s} exposure) before the current drop and from the first $38$ minutes after the drop, as shown in Figures~\ref{fig:SN9-E2-before} and \ref{fig:SN9-E2-after}. This allowed us to extract a point-like optical emission from a single APD cell, as seen in Fig.~\ref{fig:SN9-E2-diff}.

The coincidence between the \qty{\sim 0.2}{\uA} current drop and the disappearance of the emission supports the hypothesis. Additionally, the emission size that is smaller than the \qty{50}{\um} APD cell suggests that an abnormal avalanche hotspot can form in APD cells for an unknown reason.

\section{Discussion and Conclusion}
\label{sec:discussion}


The multimodal dark current shifts observed in our long-term monitoring typically range from \qty{0.1}{\uA} to \qty{0.6}{\uA}, about 10\% of the average dark current level per channel. However, the current contribution from night sky brightness is two orders of magnitude larger than this multimodal variations. The minor contribution from the unpredictable discrete shifts will not impact telescope performance.


Our two key findings---the combination of multiple current shifts (Fig.~\ref{fig:multimodal-B6}) and the correlation between avalanche emission and a current shift---suggest that multimodal SiPM dark currents originate from abnormal avalanche processes randomly occurring at specific locations in APDs.

The root cause of this behavior remains unclear from our current study. However, its similarity to the RTSs in previous studies suggests that the discrete dark current changes may be related to lattice defects in phosphor-doped silicon.

\acknowledgments

This study was supported by JSPS KAKENHI Grant Numbers JP23H04897, JP21H04468, and JP20H01916.

\pagebreak

\providecommand{\href}[2]{#2}\begingroup\raggedright\endgroup

\end{document}